\documentclass[preprint,tightenlines,showpacs,aps]{revtex4}
\usepackage{graphicx}
\begin{document}
\def\beq{\begin{equation}}
\def\eeq{\end{equation}}
\def\ds{\displaystyle}
\def\cR{{\cal R}}
\def\cI{{\cal I}}
\def\mn{_{\mu\nu}}
\def\tr{\tilde R}
\def\tt{\tilde\tau}
\def\ts{\tilde\sigma}
\def\tmu{\tilde\mu}
\def\tm{\tilde m}
\def\tmm{\tilde M}
\def\ta{\tilde\alpha}
\def\ra{\rightarrow}
\def\etal{{\it et al.\ }}

\title{The universe out of a breathing bubble}
\author{Eduardo I. Guendelman}
\email{guendel@bgu.ac.il}
\affiliation{Physics Department, Ben Gurion  University of the Negev, Beer Sheva 84105, Israel}

\author{Nobuyuki Sakai}
\email{nsakai@e.yamagata-u.ac.jp}
\affiliation{Department of Education, Yamagata University, Yamagata 990-8560, Japan}

\begin{abstract}
We consider the model of a false vacuum bubble with a thin wall where the surface energy density is composed of two different components, ``domain-wall" type and "dust" type, with opposite signs.
We find stably oscillating solutions, which we call ``breathing bubbles".
By decay to a lower mass state, such a breathing bubble could become either  i) a child universe or ii) a bubble that ``eats up" the original universe, depending on the sign of the surface energy of the ``domain-wall" component.
We also discuss the effect of the finite-thickness corrections to the thin-wall approximation and possible origins of the energy contents of our model.
\end{abstract}

\maketitle

\section{Introduction}

The inflationary Universe scenario \cite{Inflation} proposes that the observed universe originates from a region of false vacuum that expands exponentially.
Even if inflation occurred in a small region in a highly inhomogeneous spacetime, the region could evolve into the entire observable Universe.
Such a picture can resolve many cosmological puzzles, such as the horizon, flatness and monopole problems.

The detailed implementation of local inflation in an inhomogeneous spacetime has been studied in the literature. 
The original model is a spherical false-vacuum (de Sitter) bubble embedded in an asymptotically flat
(Schwarzschild) spacetime, which was devised by Sato \etal \cite{SSKM} in the context of old
inflation \cite{Inflation}. This model was investigated systematically by Blau, Guendelman, and Guth \cite{Bubbles}  with Israel`s junction conditions \cite{Israel}.
Their classical solutions show that a false-vacuum bubble can indeed expand to infinity, which is surrounded by black hole horizons and causally disconnected from the original universe.ff This new universe is called a ``child universe". 

Farhi and Guth \cite{Farhi-Guth} discussed whether such a false-vacuum bubble can be
created in the laboratory, applying the Penrose theorem \cite{Penrose}. The theorem states that, if
(a) there exists a noncompact Cauchy surface,
(b) $\cR\mn k^{\mu}k^{\nu}\ge0$ for all null vector $k^{\mu}$,
(c) there exists an anti-trapped surface,
then there exists at least one past incomplete null geodesic. As a consequence of the 
Einstein equations $G\mn=8\pi GT\mn$, condition (b) is rewritten as
$T\mn k^{\mu}k^{\nu}\ge0$. Because any standard theory of matter, 
including a canonical scalar field, obeys this energy condition, we may 
conclude that it is impossible to create an inflationary universe in the laboratory.
Condition (c) represents the realization of an inflationary universe since 
the existence of an anti-trapped surface means the existence of the cosmological horizon. 
To put it simply, a false-vacuum bubble large enough to be an inflationary 
universe cannot avoid an initial singularity, while a bubble without 
an initial singularity is too small to expand.

Several ways to avoid this obstacle have been developed. First, to escape from the Penrose theorem which applies to any classical process, Farhi, Guth, and Guven \cite{Farhi-Guth-Guven} and Fischler, Morgan, and Polchinski \cite{FMP} considered a quantum decay from a small bubble without an initial singularity 
to a large bubble which becomes an inflationary universe.
Later, Guendelman and Portnoy \cite{Guen- Port1} proposed a new model where additional matter fields are introduced to stabilize a bubble. A new aspect of this model is that an inflationary universe can be created from a stable bubble. In the $2+1$ dimensional spacetime version of that
model \cite{Guen- Port2} (where the bubble surface is a $1+1$ dimensional cord) universe creation can be achieved by an arbitrarily small tunneling, which they called ``almost classical creation of a universe".
Most recently, Sakai \etal \cite{Sakai} studied the classical and quantum creation of a universe out of a gravitating magnetic monopole.
It was shown that the classical creation of a child universe from nonsingular initial conditions could take place.
The special feature of the Reissner-Nordstr\"om spacetime that a timelike singularity exists makes it possible to create a child universe without past singularity.

It has also been argued that properties of the surface energy (tension) of a bubble can drastically affect the bubble dynamics. For example, Ansoldi and Guendelman \cite{Ansoldi-Guendelman} argued that, if the surface energy has a component that originates from stringy matter, the creation of a universe could take place at zero or arbitrarily low energy cost.
Recently Lee \etal \cite{Lee} considered a false-vacuum bubble with a nonminimally coupled scalar field. It was shown that, if the effective gravitational constant is negative, which is equivalent to the case where the surface density is negative, the bubble expands and ``eats up" the ambient universe instead of becoming a child universe.

In this paper we consider the model of a false vacuum bubble with a thin wall where the surface energy density is composed of two different components, ``domain-wall" type and ``dust" type, with opposite signs.
In this model, as we shall show below, a stable bubble in a ``breathing mode" appears; after decreasing the bubble mass, which could take place by radiation emission, the bubble starts inflation and either i) become a child universe or ii) ``eats up" the ambient universe, depending on the details of the surface energy.
Child-universe solutions in Case i) could avoid the Farhi-Guth obstacle described above, by introducing
matter which violates the energy condition.
In the simple model without finite-thickness corrections, the bubble wall temporally becomes lightlike when the surface energy vanishes; if we consider the finite-thickness corrections, however, the effect prevents it from becoming lightlike.
We shall also discuss possible origins of the exotic matter contents in Sec.\ IV.\ and possible mass loss mechanisms in Sec.\ V.

\section{Model and Basic Equations}

We consider the model of a false-vacuum bubble as follows. 
The inside is a part of de Sitter spacetime,
\beq
ds^2=-A_-dt_-^2+{dr^2\over A_-}+r^2d\Omega^2,~~~
A_-(r)\equiv1-H^2r^2,
\eeq
the outside is a part of Schwarzschild spacetime,
\beq
ds^2=-A_+dt_+^2+{dr^2\over A_+}+r^2d\Omega^2,~~~
A_+(r)\equiv1-{2GM\over r},
\eeq
and the boundary ($\Sigma$) is an infinitesimally thin matter,
\beq
S\mn\equiv\int^{+0}_{-0}T\mn dn=(\varpi+\sigma)u_{\mu}u_{\nu}+\varpi(g\mn -n_{\mu}n_{\nu})
\eeq
where $\sigma,~\varpi$, $u_{\mu}$, and $n_{\mu}$ are the surface energy density, the surface pressure, the four-velocity $\Sigma$, and the normal vector of $\Sigma$, respectively.

For matter field on $\Sigma$, we assume that the wall is composed of two components, ``domain-wall" type and ``dust" type, with opposite signs:
\beq\label{sigma}
\sigma=\mu+{m\over4\pi R^2},~~~\varpi=-\mu,~~~m\mu<0,
\eeq
where $R$ is the areal radius of the wall.
The two distinct cases i) $\mu>0,~m<0$ and ii ) $\mu<0,~m>0$ correspond to two different 
physical situations. In Case i), while $\mu>0$ represents a conventional domain wall, the ``dust" contribution $m<0$ has the opposite sign to that of ordinary dust. A simple model which can give rise to this sign could be ``ghost matter", such as an oscillating scalar field where both kinetic and mass terms
have opposite signs to the standard ones in the action. 
In Case ii) $m>0$ just represents conventional dust, and $\mu<0$ is not so exotic because it represents a negative ``cosmological constant" in the wall. In fact negative brane tensions have been considered in brane world scenarios. 

The junction condition at $\Sigma$ is given by \cite{Bubbles,Israel}
\beq\label{jc}
\beta^--\beta^+=4\pi G\sigma R,
\eeq
where
\beq\label{beta}
\beta^{\pm}\equiv\partial_nR^{\pm}
=\varepsilon^{\pm}\sqrt{\left({dR\over d\tau}\right)^2+A_{\pm}},~~~
\varepsilon=+1~{\rm or}~-1.
\eeq
Note that the sings of $\beta^{\pm}$ determine the global spacetime structure \cite{Bubbles}.
Introducing dimensionless quantities, 
\beq\label{rescale}
\tr\equiv HR,~~~ \tt\equiv H\tau,~~~
\tmm\equiv HGM,~~~ \ts\equiv{4\pi G\sigma\over H},~~~
\tmu\equiv{4\pi G\mu\over H},~~~
\tm\equiv HGm,
\eeq
we rewrite (\ref{jc}) with (\ref{beta}) as
\beq\label{eom}
\left({d\tr\over d\tt}\right)^2+U(\tr)=0,
\eeq\beq\label{U}
U(\tr)\equiv1-{\tr^2\over2}-{\tmm\over\tr}-{\ts^2\tr^2\over4}
-{1\over4\ts^2}\left(1-{2\tmm\over\tr^3}\right)^2,
\eeq\beq\label{beta2}
\beta_{\pm}={1\over2\ts\tr}\left({2\tmm\over\tr}-\tr^2\mp\ts^2\tr^2\right),~~~
\ts=\tmu+{\tm\over\tr^2}.
\eeq

Because we have assumed $\tmu\tm<0$, the sign of $\ts$ changes at 
\beq
\tr=\tr_0\equiv \sqrt{-{\tmu\over\tm}}.
\eeq
Equation (\ref{beta2}) indicates that $\beta_{\pm}$ diverges at $\tr=\tr_0$ unless
\beq
\tmm=\tmm_0\equiv{\tr_0^3\over2}=\frac12\left(-{\tm\over\tmu}\right)^{\frac32}.
\eeq
If $\tmm=\tmm_0$, the effective potential $U(\tr)$ is continuous and finite for $0<\tr<\infty$.
On the other hand, if $\tmm\ne\tmm_0$, $U(\tr)\rightarrow-\infty$ and $|dR/d\tau|\ra\infty$
as $\tr\ra\tr_0$. This means the bubble wall becomes lightlike at $\tr=\tr_0$, where the present formalism breaks down.

One way to avoid this problem is to take finite-thickness corrections into account.
According to Barrabes, Boisseau, and Sakellariadou \cite{Barrabes}, the surface energy density with the corrections are given by
\beq\label{sigma1}
\sigma_{\alpha}=\sigma_0+
{2\alpha\mu\over R^2}\left\{\left({dR\over d\tau}\right)^2+1\right\},~~~
\sigma_0\equiv\mu+{m\over4\pi R^2},
\eeq
where $\alpha$ is a correction parameter, which is of the order of (width)$^2/6$.
The basic equations (\ref{eom})-(\ref{beta2}) are unchanged except for $\sigma$.

Substituting (\ref{sigma1}) into (\ref{eom}) with (\ref{U}) and expanding it
by $\alpha/R^2$ up to the first order, we obtain the modified equation of motion,
\beq\label{eom1}
\left({d\tr\over d\tt}\right)^2+U_{\alpha}(\tr)=0,
\eeq\beq\label{U1}
U_{\alpha}(\tr)\equiv\left\{1-{\tr^2\over2}-{\tmm\over\tr}-{\ts_0^2\tr^2\over4}
-{1\over4\ts_0^2}\left(1-{2\tmm\over\tr^3}\right)^2\right\}
(1+\ta F)-\ta F,
\eeq
where
\beq
\ts_0\equiv{4\pi G\sigma_0\over H},~~~\ta\equiv\alpha H^2,~~~
F\equiv\ts_0-{1\over\ts_0^3\tr^2}\left(1-{2\tmm\over\tr^3}\right)^2.
\eeq
Although our expression (\ref{eom1}) with (\ref{U1}) look different from (3.28) with (3.29) in Ref.\cite{Barrabes}, but for the  ``dust" term, they are equivalent under the linear approximation in $\alpha$. An advantage of our form is that no higher-order term of $(d\tr/d\tt)^2$ appears; then we can discuss the bubble dynamics simply by the effective potential again.

\section{Spacetime Solutions}

\subsection{Solutions without Finite-Thickness Corrections}

Here we discuss some classical solutions of the equation of motion (\ref{eom}) with the effective potential $U(\tr)$ in (\ref{U}), which does not include the finite-thickness corrections.
We consider two cases i) $\tmu>0,~\tm<0$ and ii ) $\tmu<0,~\tm>0$; 
note that the sign transformation,  $\ts\ra-\ts$,  changes not  $U(\tr)$ (i.e., $\tr(\tt)$), but  the signs of $\beta^{\pm}$ (i.e., global spacetime structures).

\begin{figure}
\includegraphics[scale=0.5]{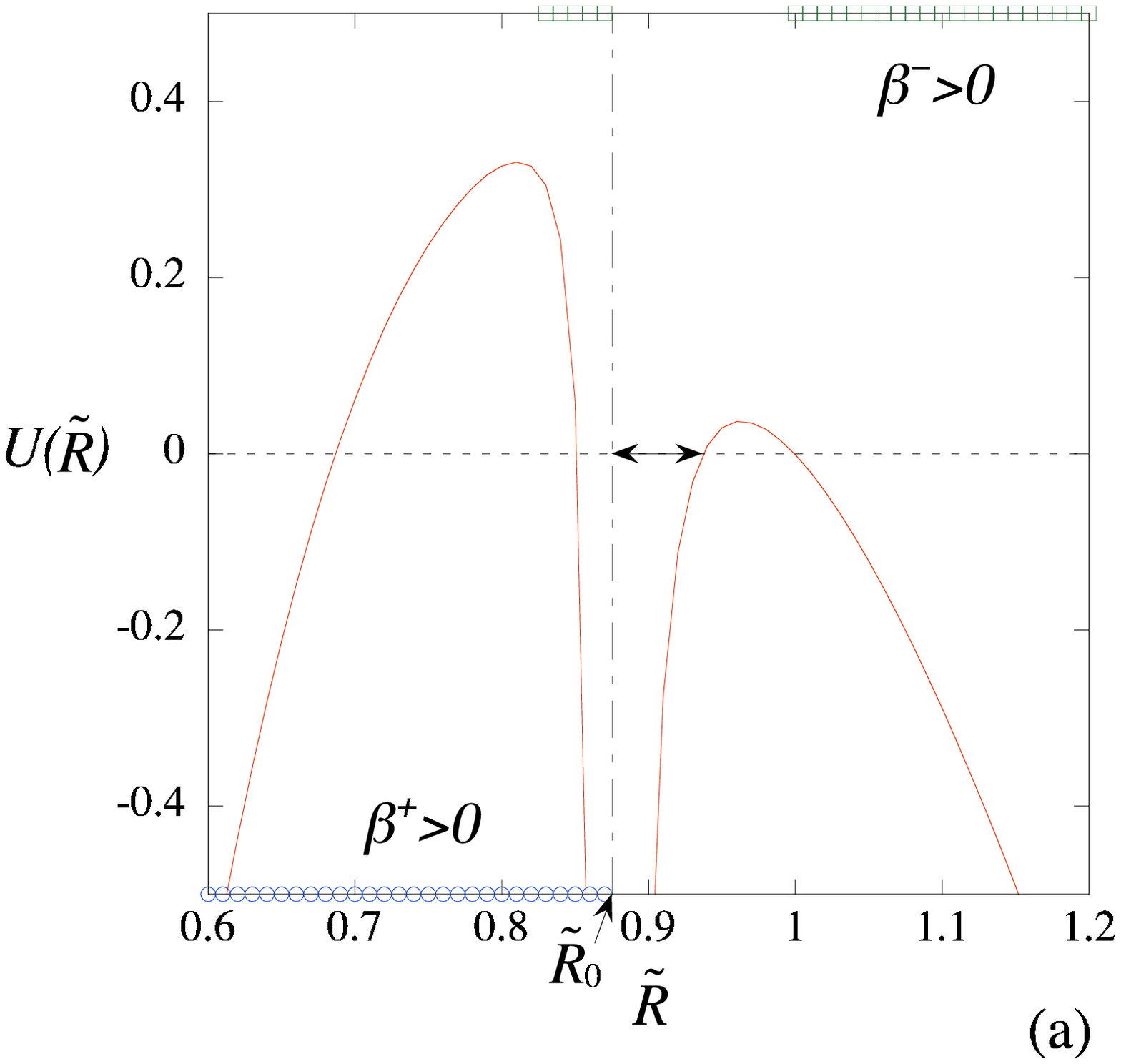}
\includegraphics[scale=0.5]{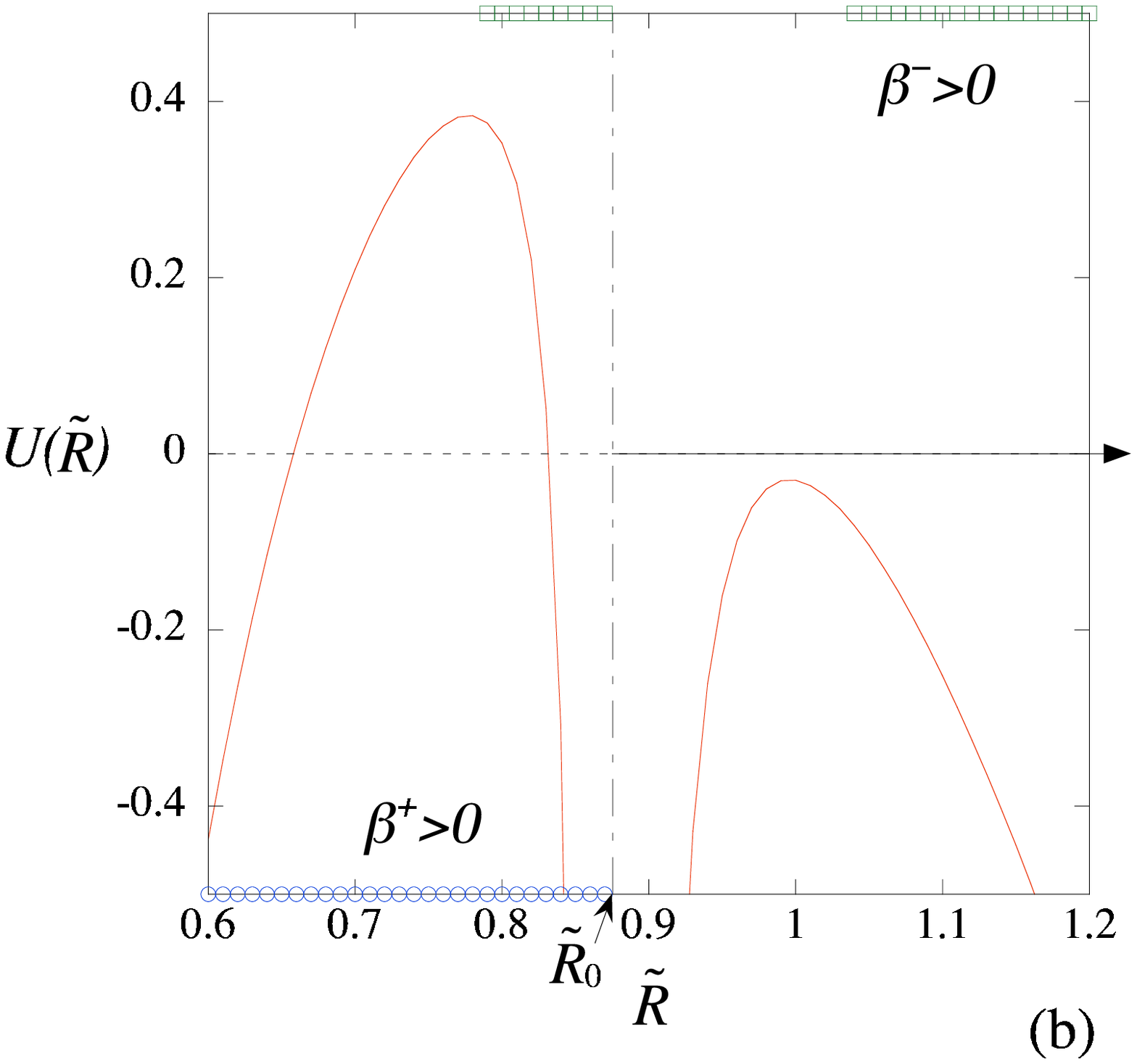}
\caption{\label{f1} Solutions without finite-thickness corrections. 
We choose $\tmu=+2,~\tm=-2.3$ and (a) $\tmm=0.7\tmm_0\approx0.23$ and (b) $\tmm=0.4\tmm_0\approx0.13$.
Circles at the top denote the region of $\beta^->0$, while squares at the bottom denote the region of $\beta^+>0$ .
(a) and (b) represent a breathing bubble and an expanding bubble, respectively.
The transition from (a) to (b) is interpreted as the creation of a child universe.}
\end{figure}

\vskip 3mm
{\bf Case i): $\mu>0,~m<0$.}

Figure 1 shows some solutions of Case i), where we depict $U(\tr)$ and the signs of $\beta^{\pm}$.
As we mentioned in Sec.\ II, $d\tr/d\tt$ and $\beta^{\pm}$ diverge at $\tr=\tr_0$.
Furthermore, because $\beta^{\pm}$ are discontinuous (
$\ds\lim_{\tr\ra\tr_0-0}\beta^{\pm}\ra+\infty$ and $\ds\lim_{\tr\ra\tr_0+0}\beta^{\pm}\ra-\infty$),
we should conclude that $\tr<\tr_0$ and $\tr>\tr_0$ correspond to different spacetime solutions, and that $\tr$ cannot cross over $\tr_0$ in either spacetime.
Although we cannot treat the system properly after the wall becomes lightlike at $\tr=\tr_0$, it is reasonable to assume that the wall reflects at $\tr=\tr_0$ and return to the timelike oscillating trajectory.
In Fig.\ 1(a), we obtain two bounded (breathing) solutions, $\tr<\tr_0$ and $\tr>\tr_0$.
Judging from the signs of $\beta^+$, the smaller bubble is located outside the Schwarzschild horizon (i.e., no horizon appears), and the larger bubble is located inside (beyond) the horizons.
Conformal diagrams of the solutions in Fig.\ 1 are shown in Fig.\ 2.

\begin{figure}
\includegraphics[scale=0.3]{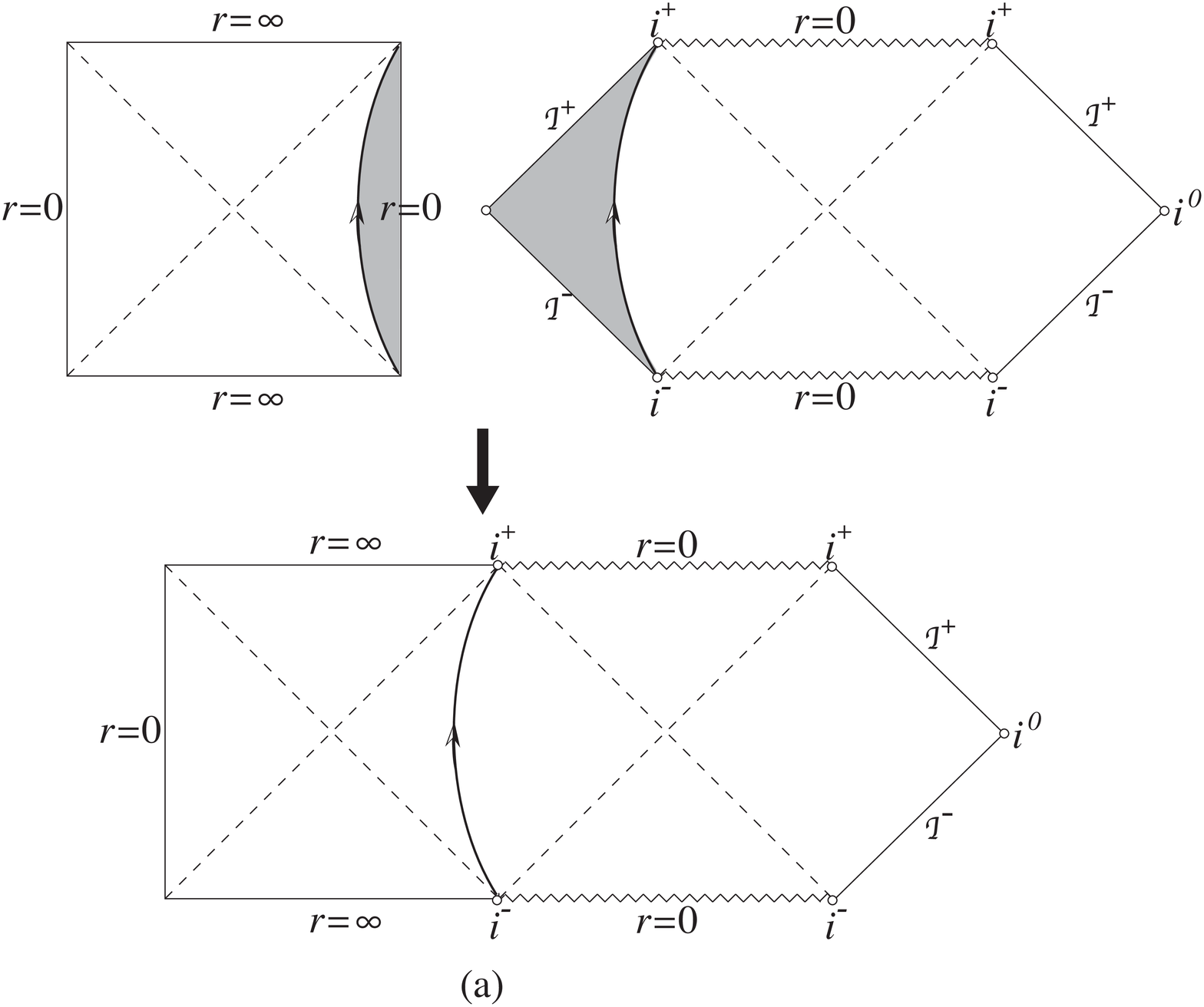}
\includegraphics[scale=0.3]{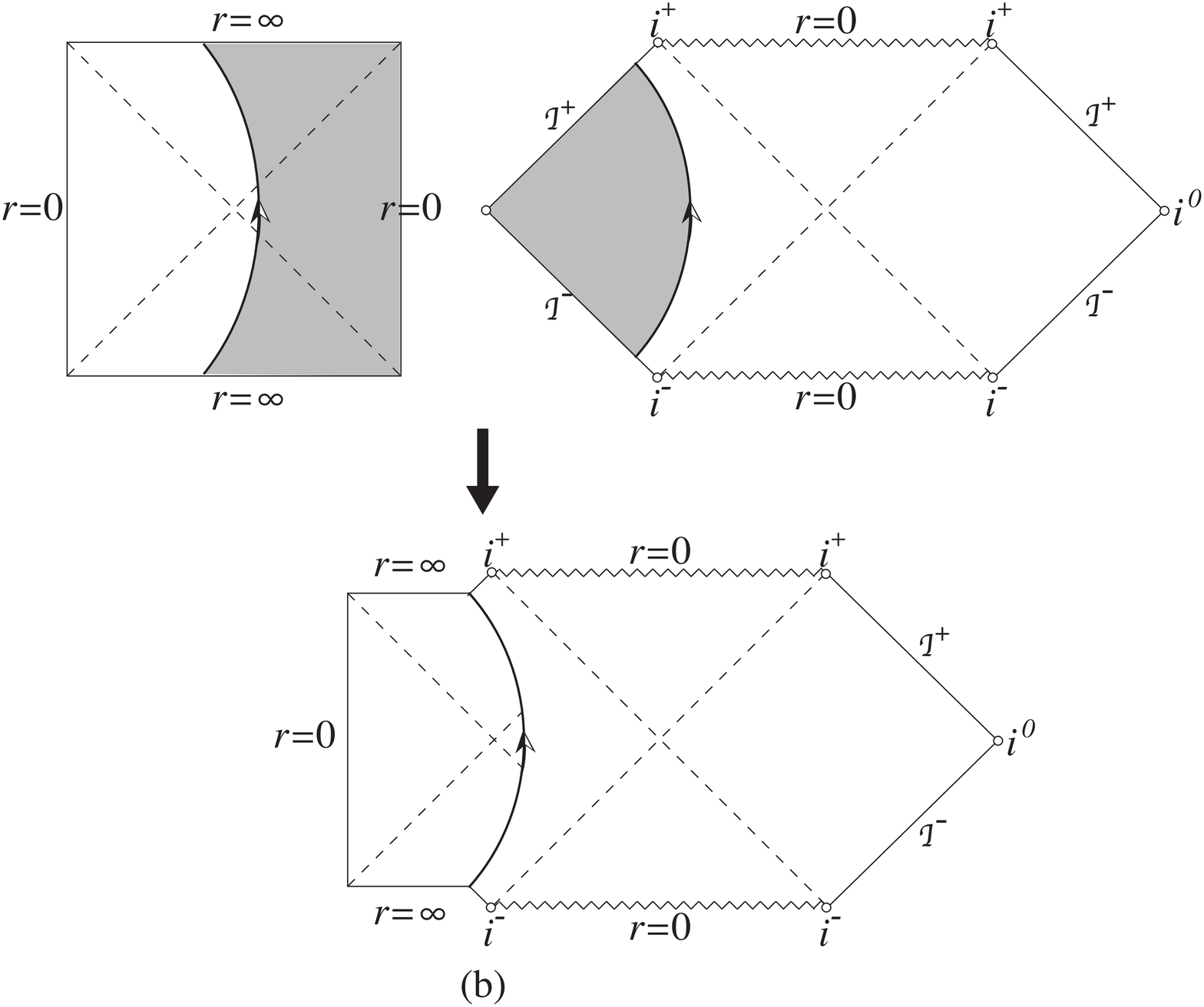}
\caption{\label{f2} Conformal diagrams of the solutions in Fig.\ 1. 
The upper figures show how the trajectory of the shell is embedded in 
Schwarzschild full spacetime and in de Sitter full spacetime, where gray domains 
indicate nonexistent regions.
The lower figures show complete spacetimes.
$\cI^+$ and $\cI^-$ represent future and past null infinity, $i^+$ and 
$i^-$ represent future and past timelike infinity, and $i^0$
represents spacelike infinity. Jaggy lines denote spacelike singularity.
Dashed lines represent Schwarzschild horizons or de Sitter horizons.}
\end{figure}

If we reduce $\tmm$ with the other parameters unchanged, we obtain an unbounded solution, where the bubble expand to infinity, as is shown in Fig.\ 1(b).
Because the sign of $\beta^+$ is negative, this expanding bubble is also located inside (beyond) the Schwarzschild horizons.
Then we expect that a breathing bubble of $\tr>\tr_0$ in (a) can evolve into an expanding bubble in (b) by mass radiation, which we shall discuss in Sec.\ V.
Because $\sigma+\varpi=m/4\pi R^2<0$, which violates the energy condition of the Penrose theorem, we could achieve the creation of a universe from nonsingular initial conditions.

This scenario is similar to that in Ref.\cite{Sakai}, where the transition from a stable magnetic monopole (described by a de Sitter bubble embedded in Reissner-Nordstr\"om spacetime) to an unbounded solution was obtained. In Ref.\cite{Sakai}, however, to realize the transition, the mass of the bubble should increase, which could be obtained by dropping matter onto the magnetic monopole. 
The process considered here is spontaneous rather than induced (like in Ref.\cite{Sakai}), since the emission of scalar 
radiation from an oscillating bubble takes place without external intervention. One would have to take care of producing a breathing bubble only.

\begin{figure}
\includegraphics[scale=0.5]{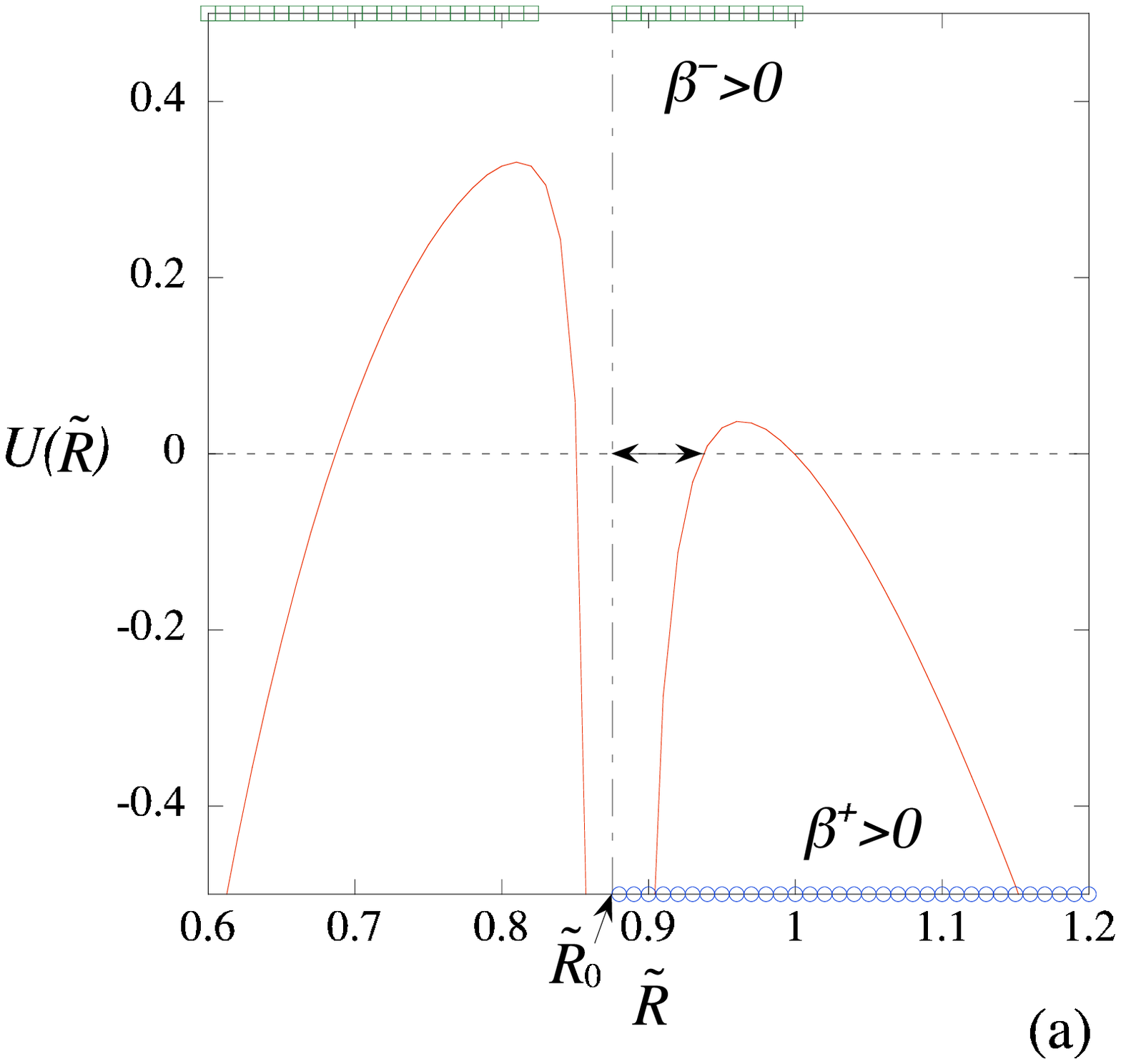}
\includegraphics[scale=0.5]{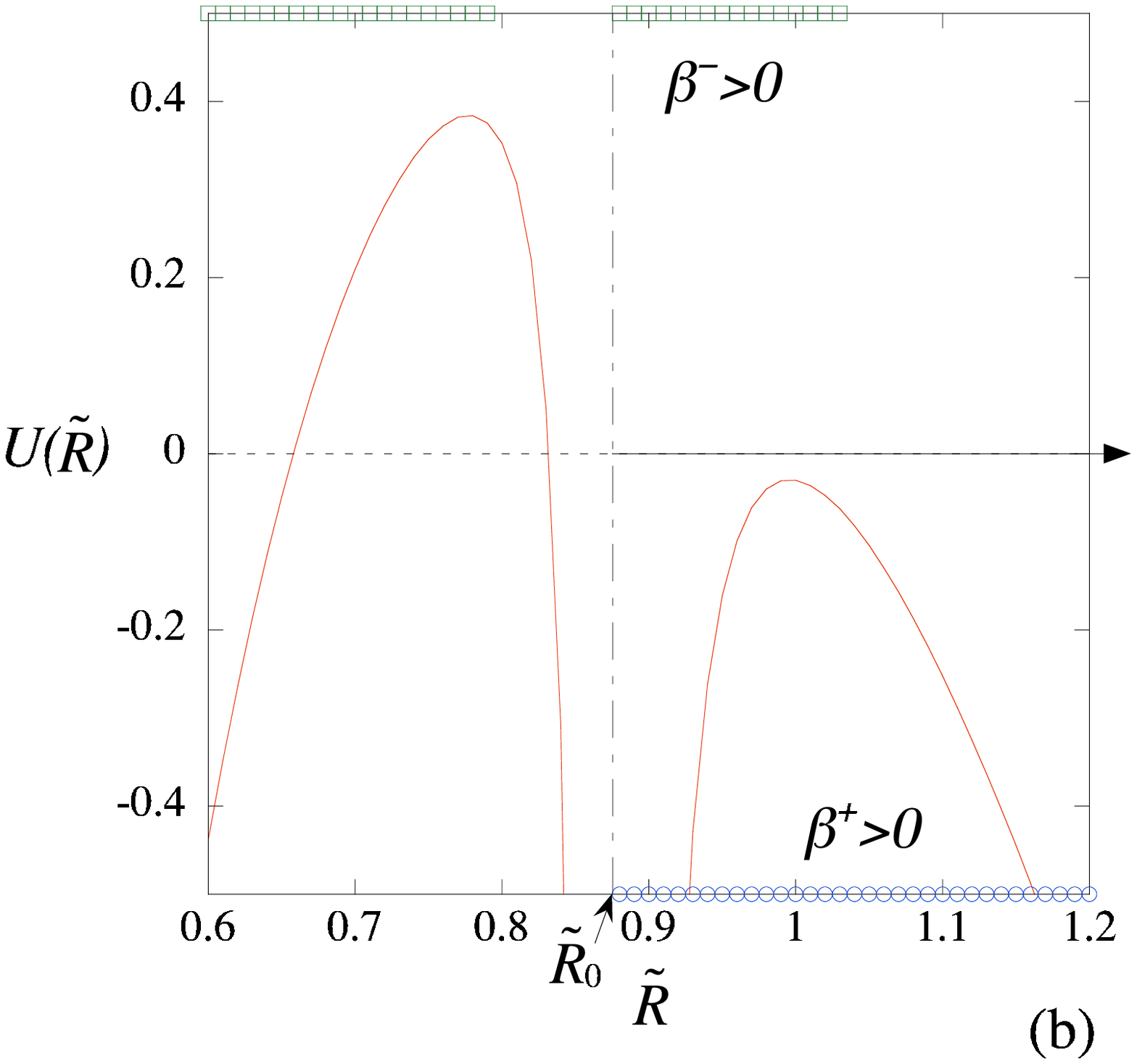}
\caption{\label{f3} Solutions without finite-thickness corrections. 
We choose $\tmu=-2,~\tm=+2.3$ and (a) $\tmm=0.7\tmm_0\approx0.23$ and (b) $\tmm=0.4\tmm_0\approx0.13$.
Although the behavior of $\tr(\tt)$ is identical to that in Fig.\ 1, the sings of $\beta^{\pm}$ are opposite, and so is the spacetime structure. (b) represents a bubble which eats up the original universe.}
\end{figure}

\vskip 3mm
{\bf Case ii): $\mu<0,~m>0$.}

Figure 3 shows some solutions of Case ii), where
the behavior of $\tr(\tt)$ is the same as that of Case i) in Fig.\ 1.
Therefore, we expect again that a breathing bubble in (a) can evolve into an expanding bubble in (b) by
a radiation process.
In contrast to Case i), however, the sign of $\beta^+$ is positive for $\tr>\tr_0$, a breathing bubble becomes not a child universe but a bubble which eats up the surrounding universe.
Conformal diagrams of the solutions in Fig.\ 3 are shown in Fig.\ 4.

\begin{figure}
\includegraphics[scale=0.3]{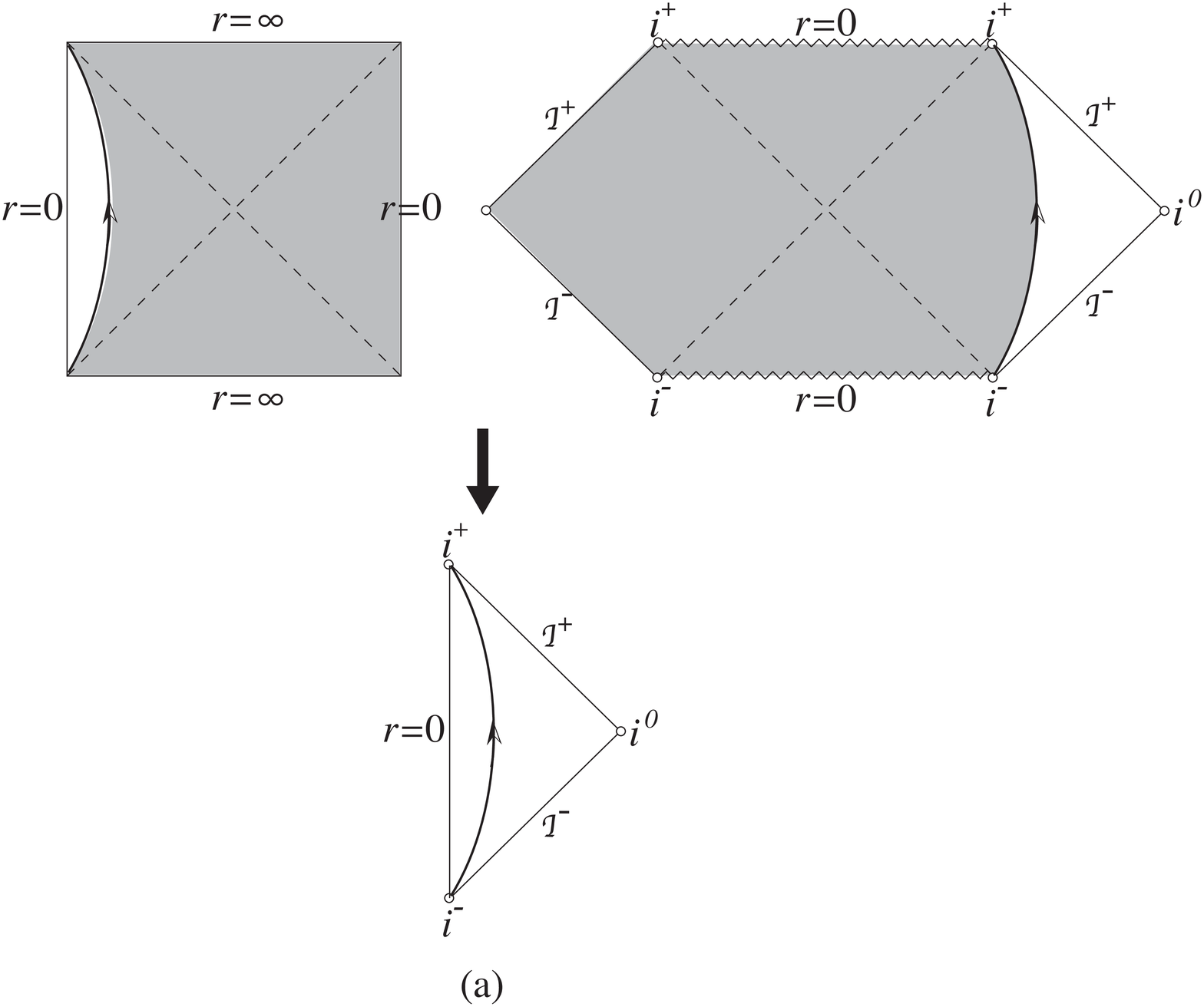}
\includegraphics[scale=0.3]{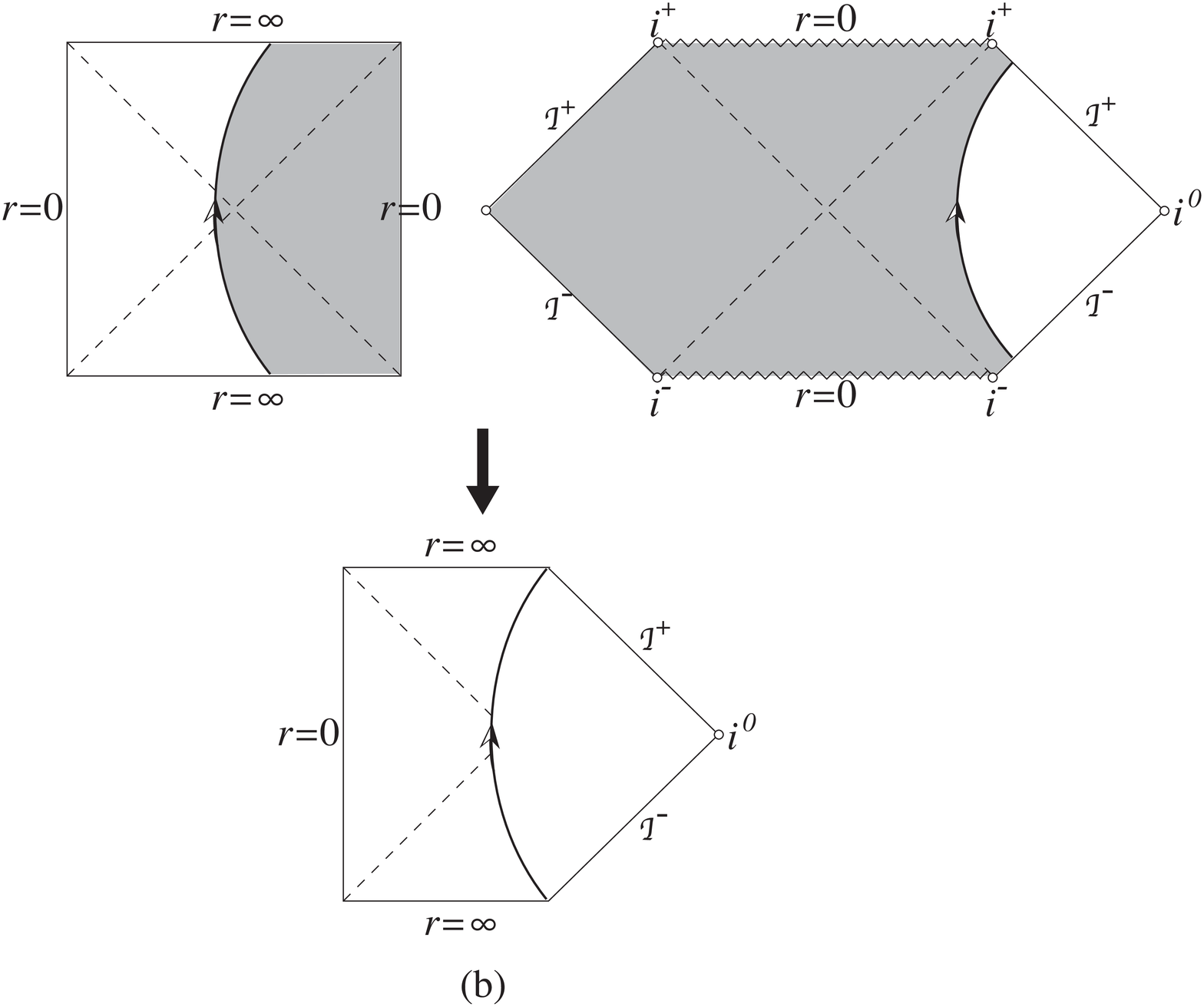}
\caption{\label{f4} Conformal diagrams of the solutions in Fig.\ 3.}
\end{figure}

\subsection{Solutions with Finite-Thickness Corrections}

\begin{figure}
\includegraphics[scale=0.5]{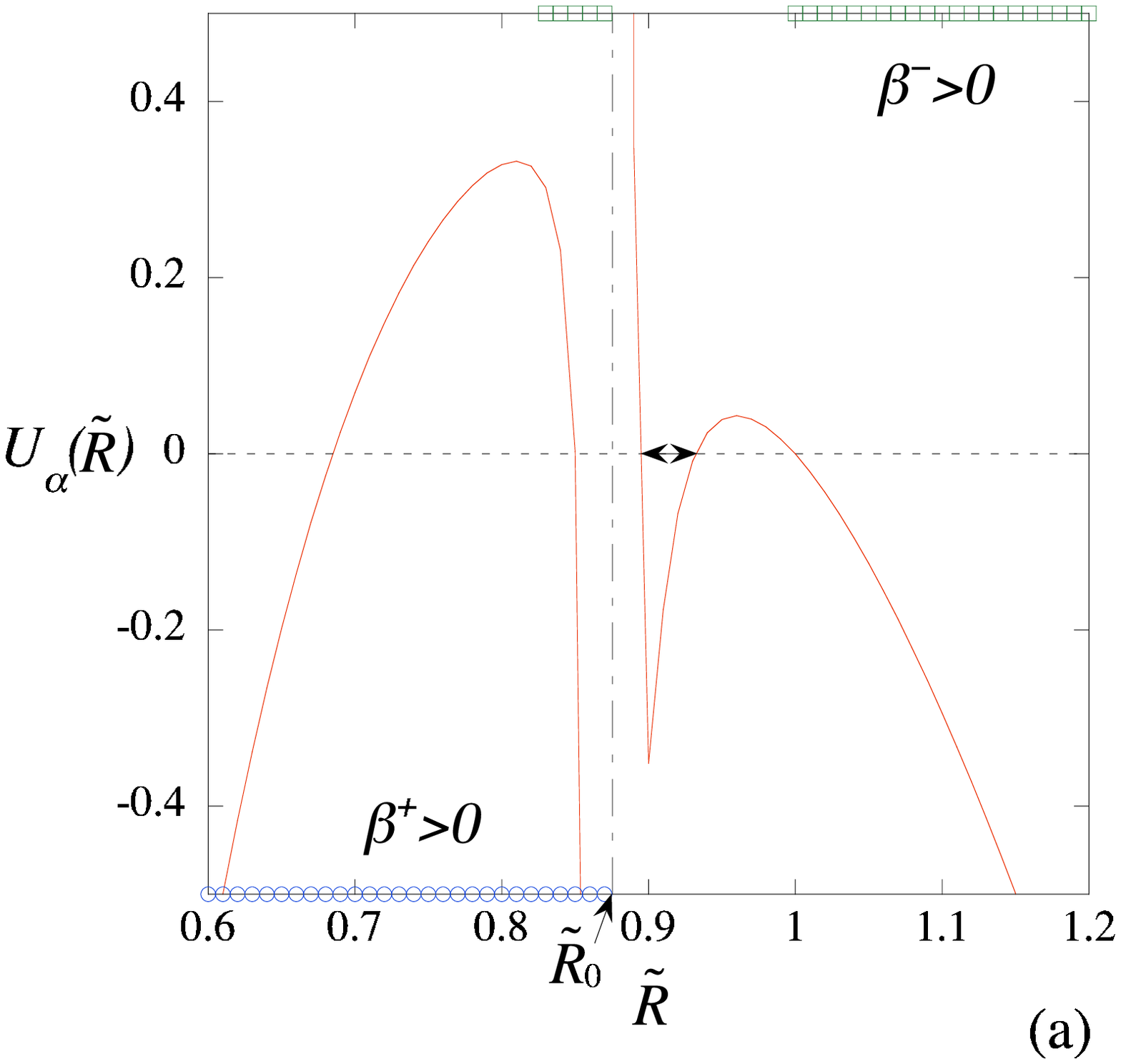}
\includegraphics[scale=0.5]{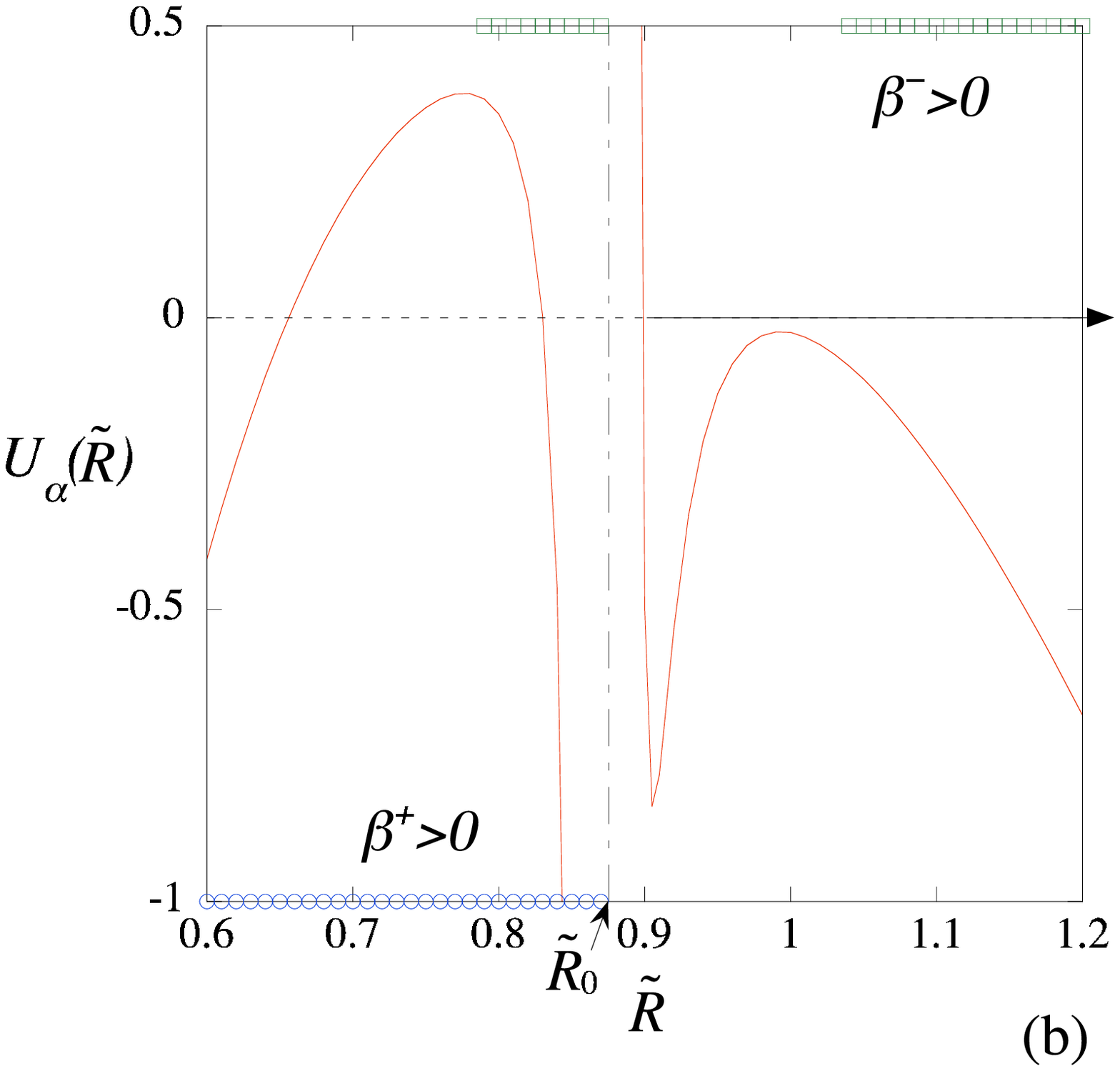}
\caption{\label{f5} Solutions with finite-thickness corrections.
We choose $\ta=0.005$;  the other parameters are the same as those in Fig.\ 1.
A barrier appears and excludes the point $\tr=\tr_0$ in a breathing solution.}
\end{figure}

In the breathing solutions discussed in Sec.\ III.A, the bubble wall becomes lightlike at $\tr=\tr_0$  ($\ts = 0)$, where the present formalism breaks down.
Although we have discussed the properties of the solutions under the reasonable assumption that the wall reflects at $\tr=\tr_0$, here we adopt an alternative way to avoid lightlike points: we consider  the corrections to the thin-wall approximation due to the finite-thickness of the domain wall, following Barrabes  \etal  \cite{Barrabes}

Figure 5 shows the effective potential (\ref{U1}) with the same parameters in Fig.\ 1 except for $\ta$.
Although there exists $\tr_0$ where the surface energy density (\ref{sigma1}) vanishes,
the effect of finite thickness generates a barrier at $\tr>\tr_0$, which excludes the problematic point $\tr=\tr_0$ in a breathing solution.
The picture described in Sec.\ III.A concerning the generation of a child universe out of a breathing bubble remains intact;
the only modification is that the lowest radius of the oscillation is now bigger.

\section{Possible Origins of Energy Contents of Thin Walls}

A crucial point from what all of our results follow is the ``exotic" nature of the energy contents of the thin wall. A special discussion of this particular point seems therefore necessary. 

\vskip 3mm
{\bf Case i): $\mu>0,~m<0$.}

We think that our use of ``negative energy dust" can be justified as at least a qualitative classical representation of quantum gravity effects.
In this respect we can mention for example the work by Padmanabhan and Narlikar \cite{Padmanabhan}, where it was found that the quantization of the conformal degrees of freedom of a metric leads to an effective energy momentum tensor with negative energy density.
Such an energy momentum tensor gives the possibility of a ``breathing  universe", very similar to a ``breathing bubble" considered here: we investigate local phenomena of a breathing bubble, while they considered global phenomena of a breathing homogeneous universe. 
Bounces have been found also in the context of loop quantum gravity \cite{Bojowald} or just by exploiting the effects of operator ordering ambiguities in the Wheeler-de Witt equation
 \cite{Guendelman-Kaganovich} to obtain a repulsive effect that prevents the universe from collapsing. 
 All of these quantum effects allow an effective classical description which however involves violation of the energy condition.
  
Yet another very interesting possibility that should be considered for producing a bounce that prevents a singularity could be the use of the ``Newton's Constant sign reversal" found by Yoshimura \cite{Yoshimura} in the high temperature Kaluza-Klein theory.
Since what appears in the junction conditions is the product of the Newton`s constant $G$ and the surface energy density $\sigma$, changing the sign of $G$ is mathematically equivalent to changing the sign of $\sigma$.
This kind of stabilization due to matter density becoming negative at some point is known to be effective in cosmology \cite{Infl-Comp}, where they can produce a bounce. Yoshimura \cite{Yoshimura} also considered bouncing universes obtained from the possible change on $G$.

The other possible origin is a gauge field which has a gradient term with the non-standard sign.
In the system of the SO(3) magnetic monopole \cite{Sakai}, the gauge field contribution to energy density is given by
$\sim(\partial_nw)/R^2$ at the wall, which has exactly the same properties as dust.
Therefore, if a gauge field has a gradient term with the non-standard sign for some reasons as discussed below, it behaves just as dust with negative energy density.
 
\vskip 3mm
{\bf Case ii): $\mu<0,~m>0$.}

A domain wall is usually defined as the localized gradient energy of a scalar field, which is given by 
$\sim(\partial_n\phi)^2$.
Therefore, if a scalar field has a gradient term with the non-standard sign, domain walls with negative energy density appear.
Such non-standard gradient (kinetic) terms may appear in string theories and have been discussed in the contest of inflation or dark energy \cite{K}.

Another possible origin is a scalar field with a standard gradient term which is localized in the 2+1 dimensional wall. We assume that the potential minimum is negative and the field oscillates around  it.
Then the kinetic energy of the field gives rise to a ``dust" contribution, while the negative value of the scalar potential at  the minimum gives rise to a ``domain-wall" contribution.
The complete system makes perfect sense even classically, so that the ``universe eating bubbles" surprisingly offer us a very simple and consistent system.
From a field-theoretical point of view, there is nothing special about supposing the potential minimum to be negative.

\section{Possible Mass Loss Mechanisms}

As we have discussed, the main mechanism which causes the transition from a breathing bubble to an infinitely expanding bubble is mass radiation.

As was shown in Ref.\cite{Sexl}, except General Relativity all alternative theories of gravity known at the time of the publication of Ref.\cite{Sexl} produce monopole radiation of the form,
\beq\label{radiation}
P= \frac{\kappa_{2}}{9\pi}\left(\frac{d^{3}A}{dt^{3}}\right)^{2},
\eeq
where $\kappa_{2}$ is a constant which is zero for GR, and $A$ is given in the linear approximation by,
\beq\label{moment}
A=\int r^{2}T_{00} d^{3}x 
\eeq
In our case we see that for non vanishing $\kappa_{2}$ we obtain a non zero radiated power and the mass loss which is required
to convert a solution that oscillates into a solution that expands to infinite radius can be achieved. Notice that $\kappa_{2}$ should be
very small, since GR is correct to a high accuracy, but then any $\kappa_{2}$ positive and non vanishing will do the job, it will be just that the required number of oscillations to achieve a given mass loss will be bigger.

Only in Case i), because Shwarzschild horizons appear, we could also expect ordinary Hawking radiation as a mass loss mechanism.

\section{Conclusions and Discussions}

We have discussed the creation of a child universe or alternatively the creation of a ``universe eating" bubble in the model where the thin wall is composed of two different components with opposite signs:
``domain-wall" type, where the surface energy density is constant,  and
``dust" type, where the surface energy density is proportional to $1/R^2$.

Because there are oscillating (breathing) bubbles as well as expanding bubbles with lower mass,
by radiation emission, a breathing bubble could be evolve into a bubble which expands to infinity.
Global spacetime structures depend on the details of the surface energy: i) if the domain-wall component is positive, this process creates a child universe, which is expanding inside (beyond) black hole horizons; ii) in the other case, no black-hole horizon appear and such a bubble just  ``eats up" the ambient universe.
Because $\sigma+\varpi=m/4\pi R^2$, the sign of the dust mass $m$ is crucial for applying the Penrose theorem. 
In Case i) of $m<0$ , we could achieve the creation of a universe from nonsingular initial conditions.

In the simple model without finite-thickness corrections, the bubble wall becomes lightlike when the total surface energy density $\sigma$ vanishes.
Although we cannot treat the lightlike boundary in the present formalism, it is reasonable to assume that the wall reflects at the moment it becomes lightlike and return to the timelike oscillating trajectory.
We also consider an alternative way to avoid lightlike points: taking the corrections to the thin-wall approximation due to the finite-thickness of the domain wall into account.
Then the oscillations are cut off at a radius slightly larger than the radius where $\sigma$ vanishes. This does not affect the above picture of the creation of a child universe from a breathing bubble by the process of loosing mass.

We have also discussed possible origins of the energy contents of thin walls. 
A negative ``dust" component in Case i) could be originated from quantization of a metric, or a gauge field with a non-standard gradient term, whereas
a negative ``domain-wall" component in Case ii) could be originated from a scalar field with a non-standard gradient term, or a scalar field with a standard gradient term which localized in the 2+1 dimensional wall.

Finally we have argued possible mass loss mechanisms.
In both cases i) and ii), the necessary mass decrease to turn breathing bubbles into expanding ones can be achieved by considering small deviations from GR that allow for monopole radiation.
Only in Case i), because Shwarzschild horizons appear, we could also expect ordinary Hawking radiation.

\acknowledgements

We thank Stefano Ansoldi and Hideki Ishihara for useful discussions.
Thanks are also due to the hospitality of Department of Physics, Osaka City University, where the substantial part of this work was carried out.
This work was supported in part by MEXT KAKENHI No.\ 18540248.


\end{document}